\documentclass[a4paper,11pt]{article}
\pdfoutput=1 

\usepackage{jcappub} 

\usepackage{latexsym,amssymb,amstext,amsmath}
\usepackage{mathrsfs}
\usepackage{enumerate}

\title{\boldmath {A Solution to the de Sitter Swampland Conjecture {\bf {\itshape {versus}}} Inflation Tension via Supergravity}}

\author{Ugur Atli}
\author{and Omer Guleryuz$^*$\note[*]{Corresponding author.}}
\affiliation{Department of Physics, Istanbul Technical University,\\Maslak 34469 Istanbul, Turkey}

\emailAdd{atliugu@itu.edu.tr}
\emailAdd{omerguleryuz@itu.edu.tr}

\abstract{The methods of supergravity allow us to derive a multi-field F-term potential. Using this, we denote a generic and non-positive single-field F-term potential. We insert our theory into the scalar-gravity part of the $SU(2,1|1)$ invariant superconformal action. That action leads us to a de Sitter solution at the inflationary trajectory. One can denote stabilization of fields in terms of the Kähler kinetic terms and single-field slow-roll inflation parameters. We combine these with the de Sitter swampland conjecture to generate a bounded conjecture. This approach allowed us to show that the single field slow-roll inflation works in harmony with bounded de Sitter conjecture for any concave inflation potential.}

\begin{document}
\maketitle
\flushbottom

	\section{Introduction}{\label{Intro}}
		
Our observable universe is in a de Sitter vacuum and, one can treat this as positive vacuum energy caused by a scalar field potential. With that motivation, constructing a de Sitter vacuum from quantum gravity is a noble quest to pursue \cite{Maloney:2002rr,Kachru:2003aw,Balasubramanian:2005zx,Westphal:2006tn,Dong:2010pm,Rummel:2011cd,Blaback:2013fca,Cicoli:2013cha,Cicoli:2015ylx,Bergshoeff:2015tra}. Considering this, one can use the suggested de Sitter swampland conjectures \cite{Obied:2018sgi,Ooguri:2018wrx} to decide whether an effective field theory is in the landscape and possibly a consistent quantum theory of gravity. These conjectures are given as:
\begin{equation}|\partial_{i} V| \geq c V\quad \text{or}\quad \min \left(\partial_{i} \partial_{j} V\right) \leq-c^{\prime} V,
\end{equation}
where $c,c' \approx \mathcal{O}(1)>0$. The first conjecture is trivial if $V$ is non-positive. Another approach to the de Sitter vacuum is the inflation theory \cite{Guth:1980zm,Starobinsky:1980te,Linde:1981mu,Albrecht:1982wi}; which leads to a de Sitter solution of gravity and suggests that the universe went through an exponential cosmic acceleration era in the early universe. This theory is well-motivated via observations \cite{Perlmutter:1998np,Aghanim:2018eyx,Percival:2009xn} and accepted as a part of the standard cosmology. So far, the single-field slow-roll inflation model is found to be in strong tension with these conjectures (see e.g. Refs. \cite{Agrawal:2018own,Achucarro:2018vey,Dias:2018ngv,Garg:2018reu,Kehagias:2018uem,Kinney:2018nny} for early seminal work). 

The main problem here; these conjectures are not fulfilled directly with a positively defined single-field inflation potential. This raises the question, can we build a model from quantum gravity candidates that lead to a positive single-field inflation potential, even though the effective potential itself is non-positive? The answer is yes. To achieve a successful model that fits in swampland conjectures and inflation theory, one can take into consideration the F-term potentials that appear in supergravity theories. In this work, we demand a de Sitter solution at the inflationary level with a generic and non-positive F-term potential. To achieve this; we can use the scalar-gravity part of the $SU(2,1|1)$ invariant superconformal action denoted in \cite{Ferrara:2010in} with an overall minus sign. This negative F-term potential allows us to discard the first conjecture since it becomes trivial. And we can denote the second conjecture together with a lower bound which, we determined from the stability conditions. Which we will refer to as \textit{the bounded de Sitter conjecture}.

This study is organized as follows. In Sec. {\ref{sec2}}, we briefly analyzed the general F-term potentials and then identified a case that corresponds to a generic non-positive F-term potential at the inflation trajectory. We also refer to the Kähler curvature and denote the required configuration to get a flat Kähler curvature at the inflationary trajectory. In Sec. {\ref{sec3}}, we show the superconformal embedding of the selected Kähler potential and the superpotential. Next, we build two single-field inflationary actions (minimally and non-minimally coupled). In Sec. {\ref{sec4}}, we get the stability conditions in terms of the Kähler kinetic terms and slow-roll parameters at the inflationary trajectory. In Sec. {\ref{sec5}}, we present the bounded de Sitter conjecture for a generic and canonical single-field slow-roll inflation (minimally coupled) model. Finally, we provide concluding remarks in Sec. {\ref{sec6}}.

\section{Generic F-term potentials in supergravity}\label{sec2}	
Since our main consideration in this work is the consistency of inflation with string derived (refined) de Sitter Swampland conjecture(s), it is natural to expect (and use) an effective potential that arises in string theory for this task. In this sense, as in the famous D-brane inflation \cite{Burgess:2001fx,Dvali:1998pa,Kachru:2003sx}, open string fields are suitable candidates for the inflationary mechanism. So, one can also use the general matter fields, which also include open strings, to obtain Kähler potentials convenient for the task like the D-brane inflation model in \cite{Kallosh:2007ig} and the fluxbrane inflation model in \cite{Hebecker:2012aw}. Following this, the embedding of the general supergravity inflation potentials \cite{Kallosh:2010xz} in a string-theoretical setup was discussed in \cite{Roest:2013aoa}. Another approach on this topic has been covered in the framework of nilpotent supergravity \cite{Dudas:2015eha,Ferrara:2014kva,Ferrara:2015tyn,Kallosh:2014via}. According to the argument in \cite{Ferrara:2014kva}, if one of the chiral superfields involved in the general inflation potentials is defined as a Nilpotent multiplet, then a simple relation can be established with the super-Dp-branes \cite{Aganagic:1996pe,Aganagic:1996nn,Bergshoeff:1997kr,Bergshoeff:1996tu,Cederwall:1996ri,Cederwall:1996pv} interacting with the $d = 10$ supergravity background. In this work, too, we will use the general string-inspired potentials with these motivations.

Starting with the assumption that the Kähler potential is separately invariant under the transformations:
\begin{align}
&S \rightarrow-S \;, \label{transformation1} \\
&\Phi \rightarrow \bar{\Phi} \;, \label{transformation2} \\
&\Phi \rightarrow \Phi+a, \quad a \in \mathbb{R} \;,  \label{transformation3}
\end{align}
where $\Phi$ and $S$ are complex scalar fields, one can derive the F-term potential in supergravity by using a Kähler potential such as:
\begin{equation}\label{GeneralK}
K(S, \bar{S},\Phi,\bar{\Phi}) \propto  S \bar{S},\; (\Phi-\bar{\Phi})^2,... \quad \xrightarrow{S=\Phi-\bar{\Phi}=0}  \quad K^{\text{IT}} = 0,
\end{equation}
which disappears at the inflationary trajectory ($S=\Phi-\bar{\Phi}=0$). Here, the scalar field $S$, serves as a stabilization term and disappears during the inflationary period. From this moment, we will use the upper letter ``IT'' to denote the inflationary trajectory $S=\Phi-\bar{\Phi}=0$, where only the real part of $\Phi$ survives. The F-term potential for $N=1$ supergravity can be written as
\begin{equation}V_F=\mathrm{e}^{\kappa^{2} K}\left(-3 \kappa^{2} W \bar{W}+\nabla_{\alpha} W g^{\alpha \bar{\beta}} \bar{\nabla}_{\bar{\beta}} \bar{W}\right),\end{equation}
where the Kähler covariant derivative is defined as $\nabla_{\alpha}W=\partial_\alpha W + \kappa^2 (\partial_\alpha K) W$ under the Kähler transformation $K \rightarrow K' = K + f + \bar{f}$ which leaves the metric $g_{\alpha \bar{\beta}}$ invariant and $\kappa \equiv M_{pl}^{-1}$. Here, $f(\Phi)$ is an arbitrary real holomorphic function, $W(\Phi)$ is the superpotential and $M_{pl}$ is the reduced Planck mass. Going one step further and defining a Kähler-invariant function \begin{equation}G=\kappa^{2} K+\log \left(\kappa^{6} W \bar{W}\right),\end{equation} helps us to write the F-term potential as it was defined in Refs. \cite{Ferrara:2016ntj, Ferrara:2019tmu}:
\begin{equation}\label{GeneralV}V_F=\kappa^{-4} \mathrm{e}^{G}\left(X-3\right),\end{equation}
where $X \equiv G_{\alpha} G^{\alpha \bar{\beta}} G_{\bar{\beta}} $ and $G_\alpha \equiv \partial_\alpha G$ etc. Since $\kappa^{2} g_{\alpha \bar{\beta}}= K_{\alpha \bar{\beta}} = G_{\alpha \bar{\beta}}$, we get $ \kappa^{-2} g^{\alpha \bar{\beta}}=K^{\alpha \bar{\beta}}=G^{\alpha \bar{\beta}} $ for inverses. We will take $\kappa = 1$ in Planckian units in the rest of this work. Using the superpotential \begin{equation}\label{W}
W=S f(\Phi),
\end{equation}
one gets:
\begin{equation}\label{eG}
\mathrm{e}^{G} = f^2 S \bar{S} \mathrm{e}^{K},
\end{equation}
and
\begin{equation}\label{X}
\begin{aligned}
X&=\frac{K^{S\bar{S}} \left(K_{SS}+1\right)\left(K_{\bar{S}\bar{S}}+1\right)}{S\bar{S}} + \frac{K^{\Phi\bar{\Phi}} \left(f K_{\Phi}+f_{\Phi}\right)\left(f K_{\bar{\Phi}}+f_{\Phi}\right)}{f^2 }  \\&+ \frac{K^{\bar{S}\Phi} \left(f K_{\Phi}+f_{\Phi}\right)\left(\bar{S} K_{\bar{S}}+1\right)}{f \bar{S} } + \frac{K^{S\bar{\Phi}} \left(f K_{\bar{\Phi}}+f_{\Phi}\right)\left(S K_{S}+1\right)}{f S } .
\end{aligned}
\end{equation}
Here, the stabilization field S belongs to the goldstino supermultiplet for this class of models. In this last equation, the composite terms $K^{S\bar{\Phi}}$ and $K^{\bar{S}\Phi}$ are always equal to zero at the inflationary trajectory for the assumed symmetries in the first place, therefore does not give any contribution to the F-term potential. Using this convention, one can derive the F-term potential at the inflationary trajectory as denoted in Ref. \cite{Kallosh:2010xz}:
\begin{equation}
V_F(0,0,\Phi,\bar{\Phi}) = \mathrm{e}^{K(0,0,\Phi,\bar{\Phi})} f(\Phi)^2 K^{S\bar{S}}(0,0,\Phi,\bar{\Phi}) \quad \xrightarrow{\Phi-\bar{\Phi}=0} \quad f(\phi)^2 K^{S\bar{S}}.
\end{equation}
One can demand the canonical normalization of the fields $\Phi$ and $S$ by arranging the Kähler potential to give a result as $K^{S\bar{S}}=K^{\Phi\bar{\Phi}}=1$ at the origin or equivalently at the inflationary trajectory. This choice results in a positive F-term potential as $ f(\Phi)^2$ and which leads to spontaneous supersymmetry breaking unless $\langle f(\Phi) \rangle =0$. In this work, our top priority is to derive a generic and non-positive F-term potential (and to test its consistency for the de Sitter swampland conjectures together with the canonical single-field slow-roll inflation). This can be achieved by using the Kähler potential
\begin{equation}\label{KählerwithNoGhost}
			    K =  -S \bar{S} + \frac{1}{2} (\Phi-\bar{\Phi})^2 + \zeta (S \bar{S})^2 + \gamma S \bar{S}(\Phi-\bar{\Phi})^2.
			\end{equation}
Note that this is a modified version of one of the Kähler potentials used in Ref. \cite{Kallosh:2010xz}.
Here, the first two terms will ensure the canonical normalization of the fields with a negative sign as $K^{S\bar{S}}=K^{\Phi\bar{\Phi}}=-1$ at the inflationary trajectory and the terms
\begin{equation}\label{second2ofK}
\zeta (S \bar{S})^2 + \gamma S \bar{S}(\Phi-\bar{\Phi})^2,
\end{equation}
will be used for the fine-tuning of the stabilization of fields. Even though those terms lead to higher-order derivatives, once the stabilization is achievable in the limit S=0 \cite{Ferrara:2010in,Lee:2010hj}, the field $S$ disappears, and the remaining field $\Phi$ determines the inflationary dynamics. It is also worth mentioning that; if one considers that the field $S$ as a nilpotent multiplet (with $S^2 = 0$), then the term $\zeta (S \bar{S})^2$ (which provides stabilization in the $S$ direction) is no longer required and also disappears. In that case, one can also eliminate the second term $\gamma S \bar{S}(\Phi-\bar{\Phi})^2$ to obtain a theory that is free from higher-order derivatives since this term is useful for fine-tuning but not essential for stabilization. Moving on, this Kähler potential respects all of the assumed symmetries (\ref{transformation1}), (\ref{transformation2}) and (\ref{transformation3}) separately. Then using this Kähler potential together with the superpotential (\ref{W}), one gets the F-term potential at the inflationary trajectory as:
\begin{equation}\label{VFIT}
V_F \xrightarrow{S=\Phi-\bar{\Phi}=0} V^{\text{IT}}_F=-f\left(\Phi\right)^2.
\end{equation}
Also the constant coefficients in front of the stabilization terms can be denoted as
\begin{equation}\label{coef}
\zeta=\frac{K_{SS\bar{S}\bar{S}}}{4},\quad \gamma=\frac{-K_{S\bar{S}\Phi\bar{\Phi}}}{2}.
\end{equation}
Furthermore, using the formula:
\begin{equation}
    R_{\text{K}} = g^{\alpha \bar{\beta}}  R_{\alpha \bar{\beta}} = -K^{\alpha \bar{\beta}} \partial_{\alpha} \left( K^{\gamma \bar{\delta}} \partial_{\bar{\beta}} K_{\gamma \bar{\delta}} \right),
\end{equation}
one can derive the curvature of selected the Kähler potential at the inflationary trajectory as:
\begin{equation}\label{flatCurv2}
    R_{\text{K}}^{\text{IT}} = -K_{SS\bar{S}\bar{S}} - 2 K_{S\bar{S}\Phi\bar{\Phi}}.
\end{equation}
One can see that the required configuration to obtain a flat Kähler curvature at the inflationary trajectory, i.e. $R_{K}^{\text{IT}} =0$, is to set the Kähler kinetic terms as $K_{SS\bar{S}\bar{S}}=- 2 K_{S\bar{S}\Phi\bar{\Phi}} $. This also shows the stabilization terms (\ref{second2ofK}) play an essential role in the flatness of the Kähler curvature.

\section{Inflation and superconformal embedding}\label{sec3}
The scalar-gravity part of the $SU(2,1|1)$ invariant superconformal action can be denoted as \cite{Ferrara:2010in}:
\begin{equation}\label{SCaction}
\mathcal{L}_{\text{sc}}^{\text{scalar-gravity}}=\sqrt{-g} \left[-\frac{1}{6} N R-G_{I \bar{J}} D_{\mu} X^{I} D^{\mu} \bar{X}^{\bar{J}}-G^{I \bar{J}} \mathcal{W}_{I} \overline{\mathcal{W}}_{\bar{J}}\right],
\end{equation}
where in this notation,  $I, \bar{J}=(0,1,2)$ symbolise to the scalar fields as $X^1\equiv \Phi$, $X^2\equiv S$ and the term $X^0$ corresponds to the first (lowest) field of the compensating multiplet. Here, all of the fields have Weyl weight 1. The superconformal Kähler potential $N(X,\bar{X})$ has Weyl weight $2$, which also satisfies the relations
\begin{equation}
N=N(X, \bar{X})=X^{I} N_{I}=\bar{X}^{I} N_{\bar{I}}=X^{I} \bar{X}^{\bar{J}} N_{I \bar{J}}.
\end{equation}
The superconformal superpotential $\mathcal{W}(X)$ has Weyl weight $3$ and satisfies
\begin{equation}\label{Wcondition}
\mathcal{W}_{I} \equiv \frac{\partial \mathcal{W}}{\partial X^{I}}, \quad \text { with } \quad X^{I} \mathcal{W}_{I}=3 \mathcal{W}.
\end{equation}
Since the kinetic term is governed by the Kähler metric, one can denote that $G_{I \bar{J}} = N_{I \bar{J}}$. For this theory, the superconformal F-term potential reads $\mathcal{V}_{F} = G^{I \bar{J}} \mathcal{W}_{I} \overline{\mathcal{W}}_{\bar{J}}$. By imposing the condition $b_{\mu} = 0$, which fixes the special conformal symmetries, the covariant derivative $D_{\mu}$ can be given as:
\begin{equation}D_{\mu} X^{I}=\left(\partial_{\mu}-\mathrm{i} A_{\mu}\right) X^{I}.\end{equation}
Here, $A_{\mu}$ is an auxiliary field of this theory and can be denoted as:
\begin{equation}A_{\mu}=\frac{\mathrm{i}}{2 N}\left(N_{\bar{I}} \partial_{\mu} \bar{X}^{\bar{I}}-N_{I} \partial_{\mu} X^{I}\right)\end{equation}
by solving the algebraic Euler-Lagrange equation for this Lagrangian. Taking consideration of the superconformal Kähler potential with Weyl weight 2,
\begin{equation}\label{SCKähler2}
    \begin{aligned}
    N=|X^0|^2 e^{\left[\frac{1}{|X^0|^2} \left(   -|X^{2}|^2 +\frac{1}{2}(X^{1}-\bar{X^{1}})^2  \right) +\frac{3 |X^{2}|^2}{|X^0|^4} \left(  \zeta |X^{2}|^2  +  \gamma  (X^{1}-\bar{X^{1}})^2 \right) \right]}
    \end{aligned}
\end{equation}
which is a modified version of one of the superconformal Kähler potentials denoted in ref. \cite{Boran:2019vxz}, one can denote the Kähler potential for the projective manifold by:
\begin{equation}\label{ProjectiveKählerPot2}
    \mathcal{K} =   -|X^{2}|^2 +\frac{1}{2}(X^{1}-\bar{X^{1}})^2   +\frac{3 |X^{2}|^2}{|X^0|^2} \left(  \zeta |X^{2}|^2  +  \gamma  (X^{1}-\bar{X^{1}})^2 \right).
\end{equation}
As one can easily see for the gauge choice $X^0 = \sqrt{3}$, which fixes the dilation symmetry, this is nothing but the Kähler potential (\ref{KählerwithNoGhost}). For this setup, the components of the superconformal Kähler metric take the values: $G_{0\bar{0}}=1$, $G_{1\bar{1}}=G_{2\bar{2}}=-1$ at the inflationary trajectory while the superconformal Kähler potential becomes: $N^{\text{IT}}=|X^0|^2$. Finally, following the condition (\ref{Wcondition}), one can denote the superconformal superpotential as:
\begin{equation}\label{SCW}
\mathcal{W}=\frac{\tau \left(X^{0}\right)^{2} X^{2} f\left(\frac{\sqrt{3}X^{1}}{X^{0}}\right)}{3} ,
\end{equation}
where $f\left(\frac{\sqrt{3} X^{1}}{X^{0}}\right)$ is an arbitrary holomorphic function and $\tau$ is real a constant. For the same gauge choice, this superconformal superpotential can be denoted as $\mathcal{W}=\tau W$ where we consider the superpotential (\ref{W}). Since we defined all the required equations and conditions, what left is to see how the action (\ref{SCaction}), evolves at the inflationary trajectory $S=\Phi-\bar{\Phi}=0$ for the superconformal Kähler potential (\ref{SCKähler2}). For this setup, the auxiliary field $A_{\mu}$ becomes zero at the inflationary trajectory, which causes that the covariant derivatives to behave as partial derivatives. Then the action (\ref{SCaction}) with an overall minus sign, leads to a physical Lagrangian with a curvature term, a canonical kinetic term and a scalar potential term as:
\begin{equation}\label{RAction}
-\mathcal{L}_{\text{sc}}^{\text{scalar-gravity}}  \xrightarrow{S=\Phi-\bar{\Phi}=0}   \mathcal{L}^{\text{IT}} =\sqrt{-g}\left[\frac{R}{2} - \left(\partial \Phi \right)^2-\tau^2 f(\Phi)^2\right].
\end{equation}
Here, the last term corresponds to the inflation potential as $V_{\text{inf}} = \tau^2 f(\Phi)^2$ with a fixed gauge $X^{0}=\sqrt{3}$. This gauge choice also allowed the curvature term in Lagrangian to appear as the standard Einstein-Hilbert term and consequently allowed us to work in the Einstein frame with a canonically normalized kinetic term. It is also worth to mention that, the relation between the superconformal F-term potential and the non-superconformal one can be given as: \begin{equation}
    \mathcal{V}_{F}^{\text{IT}}=\tau^2 V_{F}^{\text{IT}}.
\end{equation}
The F-term potential given in equation (\ref{VFIT}), can be denoted in terms of the inflation potential as:
\begin{equation}\label{VmostGeneral}
V_{F}^{\text{IT}}=-\frac{V_{\text{inf}}}{\tau^2} = - f\left(\Phi\right)^2.
\end{equation}
For the most simple case, one can pick a scalar field $\Phi$ that contains the inflaton field $\phi$ as $\Phi = \operatorname{Re} \Phi = \frac{\phi}{\sqrt{2}}$ at the inflationary trajectory. Using this F-term potential as the effective potential of our theory will grant us great freedom because this potential is allowed to be negative\footnote{A negative effective potential, easily satisfies the de Sitter swampland conjectures as given in \cite{Obied:2018sgi,Ooguri:2018wrx}.}, even though the inflation potential is positively defined. Here, we denote the superconformal Lagrangian with an overall minus sign in order to achieve a theory that leads to a de Sitter solution\footnote{In a scenario where the field $\Phi$ is a constant, the potential term can be treated as a constant too. Assuming that, $V_{\text{inf}} \rightarrow \Lambda$ is the cosmological constant, this theory leads to a de Sitter solution as:
$
a(t)=\mathrm{e}^{Ht}=\mathrm{e}^{\sqrt{\Lambda /3} t}
$.
} in the end. So far, we build a single field inflationary Lagrangian with a standard Einstein-Hilbert term
\begin{equation}
\frac{1}{\sqrt{-g}} \mathcal{L}_{\text{EH}} = \frac{1}{2}R
,\end{equation}
as $ \mathcal{L}^{\text{IT}}= \mathcal{L}_{\text{EH}} +  \mathcal{L}_{\text{matter}}$. To build a non-minimally coupled (to curvature $R$) inflation Lagrangian, one needs to adjust the superconformal Kähler potential (\ref{SCKähler2}) by introducing additional terms that have Weyl weight $2$. In this sense, one can denote:
\begin{equation}
    N' \rightarrow N + \frac{3 \xi \mathcal{W}}{\tau X^2} + \frac{3 \xi \bar{\mathcal{W}}}{\tau \bar{ X^2}},
\end{equation}
where $\mathcal{W}$ is the same superconformal superpotential (\ref{SCW}) and $\xi$ is a real constant. These additional terms, preserves the superconformal F-term potential $\mathcal{V}_{F}^{\text{IT}} = -\tau^2 f(\Phi)^2$. 
The resulting Jordan frame Lagrangian comes with a non-minimal coupling term that is proportional to an arbitrary function $f(\Phi)$ as:
\begin{equation}\label{nonminimalAction}
   -\mathcal{L}_{\text{sc}}^{\text{scalar-gravity}}  \xrightarrow{S=\Phi-\bar{\Phi}=0} \mathcal{L}^{\text{IT}}_{\text{non-minimal}} =\sqrt{-g}\left[\frac{R}{2} +\xi f(\Phi) R - \left(\partial \Phi \right)^2-\tau^2 f(\Phi)^2\right].
\end{equation}
Here, one can use the parameter $\xi$ as a coupling constant with positive or negative real values. Moreover, equation of motion of the field $\Phi$:
\begin{equation}
     f_{\Phi}\left( \xi R-2 \tau^2 f  \right) - 2\Box{\Phi} =0,
\end{equation}
leads to an effective $R+R^2$ gravity in the quasi-static limit ($\Phi \simeq \text{constant}$) as:
\begin{equation}\label{R+R2}
    \mathcal{L}= \sqrt{-g}\left[\frac{ R}{2}+ \frac{\xi^2 R^2}{4 \tau^2 }\right].
\end{equation}
As shown in ref. \cite{Kehagias:2013mya}, the quasi-static limit does not disturb the inflationary predictions because the contribution of the kinetic terms is negligible compared to the contribution of the potential terms during inflation period. In fact this limit only leads an order of $10^{-3}$ change in the spectral index $n_s$. 

The Lagrangian (\ref{nonminimalAction}) possess a generalized non-minimal coupling term $\xi f(\Phi)$, and which is proportional to the square root of the generalized inflation potential $\tau^2 f(\Phi)^2$. This kind of non-minimal coupling has been widely studied in the aspects of inflation \cite{Barbon:2009ya,Chakravarty:2013eqa,Kaiser:2013sna,Kallosh:2013tua,Pallis:2010wt,Qiu:2012ia}. In particular, ref. \cite{Kallosh:2013tua} examines this model with a similar Kähler potential using the superconformal approach and obtained a universal attractor behavior at the strong coupling limit (for large values of $\xi$). Independently of the original scalar potential, that universal attractor shares the inflationary predictions of the Starobinsky model $R+R^2$ \cite{Starobinsky:1980te}, which one can expect for the model (\ref{R+R2}).

\section{Stability conditions at the inflationary trajectory}\label{sec4}
The characteristics of the mass (squared) matrix are determined uniquely for each Kähler potential and superpotential that one considers. In this section, we will consider the stability conditions for the case that eventually points to the single-field slow-roll inflation model (\ref{RAction}).

The scalar fields $S$ and $\Phi$ can be manipulated as desired for the situation, in this case we choose the representation as:
\begin{equation}S=\frac{1}{\sqrt{2}}(s+i \alpha), \quad \Phi=\frac{1}{\sqrt{2}}(\phi+i \beta),\end{equation}
to switch the real canonically normalized fields $s,\alpha,\phi,\beta$. From now on, we will assume the field $\phi$ as the inflaton field. Then using the supergravity F-term potential formula (\ref{GeneralV}), one can derive the squared mass terms $m_{s}^{2}=\partial^{2}_{s} V$ etc. at the inflationary trajectory $s=\alpha=\beta=0$ (or $S=\Phi -\bar{\Phi} =0$). This will allow us to study stability conditions through the Kähler kinetic terms. By applying this procedure, the diagonal terms of mass matrix of all real fields are denoted as: 
\begin{equation}
\begin{aligned}
m_{s}^{2} &= m_{\alpha}^{2} = -f^2 K_{SS\bar{S}\bar{S}}  - (f_{\phi})^2, \\ m_{\beta}^{2} &=f^2 \left( -2K_{S\bar{S}\Phi\bar{\Phi}} + 2 \right) +f f_{\phi \phi} -(f_{\phi})^2, \\ m_{\phi}^{2} &=-f f_{\phi \phi} -(f_{\phi})^2,
\end{aligned}
\end{equation}
where all the terms and derivatives are calculated at the inflationary trajectory. For this case, the off-diagonal entities of the mass matrix are equivalent to zero. One can write this equations by using the inflation potential such as:
\begin{equation}V_{\text{inf}}= \tau^2 f^{2}  \simeq 3 H^{2},\end{equation}
which leads to the slow-roll parameters as:
\begin{equation}\label{slow-roll-parameters}
\left(\frac{d_{\phi} V_{\text{inf}}}{V_{\text{inf}}}\right)^{2} = \left(\frac{f_\Phi}{f}\right)^{2}=\epsilon, \quad \frac{d_{\phi}^{2} V_{\text{inf}}}{V_{\text{inf}}}  = \frac{f_{\Phi \Phi}}{f} +\epsilon=\eta.
\end{equation}
Then the diagonal entities of the mass matrix reads:
\begin{equation}\label{massTermsDiagonal}
\begin{aligned}
m_{s}^{2} &= m_{\alpha}^{2} = \frac{3 H^2}{\tau^2}\left(-K_{SS\bar{S}\bar{S}}-\epsilon\right), \\ m_{\beta}^{2} &=\frac{3 H^2}{\tau^2}\left( -2K_{S\bar{S}\Phi\bar{\Phi}} +2 -2\epsilon + \eta \right), \\ m_{\phi}^{2} &=-\frac{3 H^2}{\tau^2} \eta.
\end{aligned}
\end{equation}
Assuming that, we have a standard single-field slow-roll inflationary scenario which requires $ 0 \lesssim m_{\phi}^{2}$ for a light inflaton field together with the condition $H^2 \lesssim m_{s}^{2},m_{\alpha}^{2},m_{\beta}^{2}$ \cite{Kallosh:2010xz,Ferrara:2010yw,Ferrara:2010in}, leads us to write the stability conditions:
\begin{equation}\label{StabilityConditions}
\begin{aligned}
\frac{\tau^2}{3} \lesssim& -K_{SS\bar{S}\bar{S}} -\epsilon ,\\ \frac{\tau^2}{3} \lesssim &  -2K_{S\bar{S}\Phi\bar{\Phi}}+2  -2\epsilon + \eta , \\ 0 \lesssim& -\eta.
\end{aligned}
\end{equation}
One can immediately see from the third condition that $\eta$ needs to be negative to fulfill this condition which also indicates that the inflation potential must be concave $(d_{\phi}^{2}V_{\text{inf}} = d^2 V_{\text{inf}}/d \phi^2 < 0)$. During the slow-roll approximations $\epsilon,|\eta| \ll 1$, first two conditions obey:
\begin{equation}\label{stabilitycond}
K_{SS\bar{S}\bar{S}}  \lesssim -\frac{\tau^2}{3}, \quad K_{S\bar{S}\Phi\bar{\Phi}} \lesssim 1 - \frac{\tau^2}{6} .
\end{equation}
As one can see, there is only one stabilization branch for this case. These conditions also show that there is an achievable stability region for the flat Kähler curvature case, $R_{K}^{\text{IT}}=0$.

\section{Bounded de Sitter conjecture}\label{sec5}
The refined de Sitter conjectures \cite{Obied:2018sgi,Ooguri:2018wrx} states that an effective field theory (EFT) in the landscape, should fulfill either,
\begin{equation}|\partial_{i} V| \geq c V, \quad c \approx \mathcal{O}(1)>0,\end{equation}
or
\begin{equation} \min \left(\partial_{i} \partial_{j} V\right) \leq-c^{\prime} V, \quad c^{\prime} \approx \mathcal{O}(1)>0,\end{equation}
where $\min \left(\partial_{i} \partial_{j} V\right)$ denotes the minimum eigenvalue of $ \partial_{i} \partial_{j} V$. In this notation $i,j=(\phi,\beta,s,\alpha)$ correspond to real canonically normalized fields. For any negative F-term potential $V_F<0$, the first form of the conjecture becomes trivial. For that matter, we can use the relation $ \tau^2 V_{F}^{\text{IT}}(\phi/\sqrt{2}) = - V_{\text{inf}}(\phi/\sqrt{2})$ as denoted in the equation (\ref{VmostGeneral}), which leads to the models given in (\ref{RAction}) and (or) (\ref{nonminimalAction}) with possible de Sitter solutions.\footnote{Here, we will consider the model (\ref{RAction}) which is in the Einstein frame.} This shows that using the F-term potential (\ref{VmostGeneral}) as the effective potential provides substantial control for determining an EFT about whether it is in the landscape or swampland. As discussed in the previous section for the stabilization of fields, the properties of the remaining de Sitter conjecture are also uniquely determined for each different Kähler potential and superpotential. Then, for the second de Sitter conjecture, we need to denote the minimum eigenvalue of the term $\partial_{i} \partial_{j} V$ or equivalently minimum eigenvalue of the mass matrix. Since off-diagonal terms of the mass matrix are equivalent to zero for this case, one can determine the second de Sitter conjecture directly from the diagonal terms. Using the equation (\ref{massTermsDiagonal}) one can conclude that the minimum eigenvalue can be determined as $ m_{\phi}^{2} =-\frac{3 H^2}{\tau^2} \eta$ since each entity of this mass matrix corresponds to an eigenvalue. This allows us to denote the bounded de Sitter conjecture as:
\begin{equation}
    0 \lesssim - \eta \leq c',
\end{equation}
with a lower bound caused by the stability condition given in equations (\ref{StabilityConditions}). The highest value of  $ - \eta $ can be determined by single-field slow-roll inflation models that are observationally consistent for a concave potential. Then this inequality can also be used to determine the lowest possible value of the parameter $ c '$, which can be much smaller than $\mathcal{O}(1)$.

It is also clarifying to mention that; for a positive and generic F-term potential that driven under the transformations (\ref{transformation1}), (\ref{transformation2}) and (\ref{transformation3}) would lead to that conjecture as $0 \lesssim \eta \leq - c'$. For this case, the bounded conjecture is impossible to fulfill because of the lower bound that comes from the stability condition.

\section{Conclusions}\label{sec6}
		In this study, we revisited the de Sitter swampland conjectures by distinguishing the effective potential and inflationary potential from superconformal embedding of the superconformal action (\ref{SCaction}) with an overall minus sign. After the gauge fixing the dilation symmetry, the minimally coupled single-field inflationary action appeared as:
		\begin{equation}\label{ConclusionsRAction}
\mathcal{L}^{\text{IT}} =\sqrt{-g}\left[\frac{R}{2} - \left(\partial \Phi \right)^2-\tau^2 f(\Phi)^2\right],
\end{equation}
        and we determined the F-term potential in terms of the inflation potential as:
\begin{equation}\label{ConclusionsVmostGeneral}
V_{F}^{\text{IT}}=-\frac{V_{\text{inf}}}{\tau^2} = - f\left(\Phi\right)^2.
\end{equation}
        We also briefly analyzed a non-minimally coupled action setup at the superconformal level which eventually led us to an effective $ R + R ^ 2 $ gravity model in a quasi-static limit at the inflationary trajectory. Furthermore, we denoted the necessary stability conditions to bound the refined de Sitter conjecture for the canonical single-field slow-roll inflation scenario with a generic inflation potential. By doing so, we showed that the stabilization of the real scalar field $\phi$ (inflaton) required the condition $0 \lesssim -\eta$, which forced that inflation potential to become a concave potential (i.e. $d_{\phi}^{2}V_{\text{inf}} = d^2 V_{\text{inf}}/d \phi^2 < 0$) .
Finally, we denoted the bounded de Sitter conjecture \begin{equation}
    0 \lesssim - \eta \leq c',
\end{equation}
which set a lower bound to the parameter $c'$ with the slow-roll parameter $-\eta$. In particular, we showed that the canonical single-field slow-roll inflation mechanism that appeared from the superconformal action works in harmony with the bounded de Sitter conjecture for any concave inflation potential as long as $-\eta \leq c' \approx \mathcal{O}(1)$. 

\acknowledgments

The authors would like to thank Mehmet Ozkan for his comments and insightful suggestions. The work of Ugur Atli is supported by TUBITAK grant 118F091.



\begin{thebibliography}{99}

\bibitem{Maloney:2002rr}
Alexander Maloney, Eva Silverstein, and Andrew Strominger,
\newblock {\em De Sitter space in noncritical string theory},
\newblock in {\em {The future of theoretical physics and cosmology: Celebrating
Stephen Hawking’s 60th birthday. Proceedings, Workshop and Symposium,
Cambridge, UK, January 7-10, 2002}},  pages 570-591.
\newblock {[\href{https://arxiv.org/abs/hep-th/0205316}{hep-th/0205316}]}.



			
\bibitem{Kachru:2003aw}
Shamit Kachru, Renata Kallosh, Andrei~D. Linde, and Sandip~P. Trivedi,
\newblock {\em De Sitter vacua in string theory},
\newblock \href{ https://doi.org/10.1103/PhysRevD.68.046005}{ {\em Phys. Rev. D} 68 (2003) 046005}
\newblock {[\href{https://arxiv.org/abs/hep-th/0301240}{hep-th/0301240}]}.



			
\bibitem{Balasubramanian:2005zx}
Vijay Balasubramanian, Per Berglund, Joseph~P. Conlon, and Fernando Quevedo,
\newblock {\em Systematics of moduli stabilisation in Calabi-Yau flux
  compactifications},
\newblock \href{ https://doi.org/10.1088/1126-6708/2005/03/007}{ {\em JHEP} 03 (2005) 007}
\newblock {[\href{https://arxiv.org/abs/hep-th/0502058}{hep-th/0502058}]}.




\bibitem{Westphal:2006tn}
Alexander Westphal,
\newblock {\em de Sitter string vacua from Kähler uplifting},
\newblock \href{ https://doi.org/10.1088/1126-6708/2007/03/102}{ {\em JHEP} 03 (2007) 102}
\newblock {[\href{https://arxiv.org/abs/hep-th/0611332}{hep-th/0611332}]}.





\bibitem{Dong:2010pm}
Xi~Dong, Bart Horn, Eva Silverstein, and Gonzalo Torroba,
\newblock {\em Micromanaging de Sitter holography},
\newblock \href{ https://doi.org/10.1088/0264-9381/27/24/245020}{ {\em Class. Quant. Grav.} 27 (2010) 245020}
\newblock {[\href{https://arxiv.org/abs/1005.5403}{1005.5403}]}.




\bibitem{Rummel:2011cd}
Markus Rummel and Alexander Westphal,
\newblock {\em A sufficient condition for de Sitter vacua in type IIB string
  theory},
\newblock \href{ https://doi.org/10.1007/JHEP01(2012)020}{ {\em JHEP} 01 (2012) 020}
\newblock {[\href{https://arxiv.org/abs/1107.2115}{1107.2115}]}.





\bibitem{Blaback:2013fca}
Johan Bl\r{a}b\"ack, Ulf Danielsson, and Giuseppe Dibitetto,
\newblock {\em Accelerated Universes from type IIA Compactifications},
\newblock \href{ https://doi.org/10.1088/1475-7516/2014/03/003}{ {\em JCAP} 03 (2014) 003}
\newblock {[\href{https://arxiv.org/abs/1310.8300}{1310.8300}]}.





\bibitem{Cicoli:2013cha}
Michele Cicoli, Denis Klevers, Sven Krippendorf, Christoph Mayrhofer, Fernando
  Quevedo, and Roberto Valandro,
\newblock {\em Explicit de Sitter Flux Vacua for Global String Models with Chiral
  Matter},
\newblock \href{ https://doi.org/10.1007/JHEP05(2014)001}{ {\em JHEP} 05 (2014) 001}
\newblock {[\href{https://arxiv.org/abs/1312.0014}{1312.0014}]}.




\bibitem{Cicoli:2015ylx}
Michele Cicoli, Fernando Quevedo, and Roberto Valandro,
\newblock {\em De Sitter from T-branes},
\newblock \href{ https://doi.org/10.1007/JHEP03(2016)141}{ {\em JHEP} 03 (2016) 141}
\newblock {[\href{https://arxiv.org/abs/1512.04558}{1512.04558}]}.




\bibitem{Bergshoeff:2015tra}
Eric~A. Bergshoeff, Daniel~Z. Freedman, Renata Kallosh, and Antoine
  Van~Proeyen,
\newblock {\em Pure de Sitter Supergravity},
\newblock \href{ https://doi.org/10.1103/PhysRevD.92.085040}{ {\em Phys. Rev. D} 92(8) (2015) 085040},
\newblock \href{ https://doi.org/10.1103/PhysRevD.93.069901}{ Erratum:{\em Phys. Rev. D} 93 (2016) 069901}
\newblock {[\href{https://arxiv.org/abs/1507.08264}{1507.08264}]}.






\bibitem{Obied:2018sgi}
Georges Obied, Hirosi Ooguri, Lev Spodyneiko, and Cumrun Vafa,
\newblock {\em De Sitter Space and the Swampland}
\newblock (2018)
\newblock {[\href{https://arxiv.org/abs/1806.08362}{1806.08362}]}.





\bibitem{Ooguri:2018wrx}
Hirosi Ooguri, Eran Palti, Gary Shiu, and Cumrun Vafa,
\newblock {\em Distance and de Sitter Conjectures on the Swampland},
\newblock \href{ https://doi.org/10.1016/j.physletb.2018.11.018}{ {\em Phys. Lett. B} 788 (2019) 180-184}
\newblock {[\href{https://arxiv.org/abs/1810.05506}{1810.05506}]}.




\bibitem{Albrecht:1982wi}
Andreas Albrecht and Paul~J. Steinhardt,
\newblock {\em Cosmology for Grand Unified Theories with Radiatively Induced
  Symmetry Breaking},
\newblock \href{ https://doi.org/10.1103/PhysRevLett.48.1220}{ {\em Adv. Ser. Astrophys. Cosmol.} 3 (1987) 158-161}.




\bibitem{Guth:1980zm}
Alan~H. Guth,
\newblock {\em The Inflationary Universe: A Possible Solution to the Horizon and
  Flatness Problems},
\newblock \href{ https://doi.org/10.1103/PhysRevD.23.347}{ {\em Adv. Ser. Astrophys. Cosmol.} 3 (1987) 139-148}.



\bibitem{Linde:1981mu}
Andrei~D. Linde,
\newblock {\em A New Inflationary Universe Scenario: A Possible Solution of the
  Horizon, Flatness, Homogeneity, Isotropy and Primordial Monopole Problems},
\newblock \href{ https://doi.org/10.1016/0370-2693(82)91219-9}{ {\em Adv. Ser. Astrophys. Cosmol.} 3 (1987) 149-153}.





\bibitem{Starobinsky:1980te}
Alexei~A. Starobinsky.
\newblock {\em A New Type of Isotropic Cosmological Models Without Singularity},
\newblock \href{ https://doi.org/10.1016/0370-2693(80)90670-X}{ {\em Adv. Ser. Astrophys. Cosmol.} 3 (1987) 130-133}.




\bibitem{Aghanim:2018eyx}
PLANCK Collaboration, N.~Aghanim et~al.,
\newblock {\em Planck 2018 results. VI. Cosmological parameters},
\newblock \href{ https://doi.org/10.1051/0004-6361/201833910}{ {\em Astron. Astrophys.} 641 (2020) A6}
\newblock {[\href{https://arxiv.org/abs/1807.06209}{1807.06209}]}.





\bibitem{Percival:2009xn}
Will~J. Percival et~al.,
\newblock {\em Baryon Acoustic Oscillations in the Sloan Digital Sky Survey Data
  Release 7 Galaxy Sample},
\newblock \href{ https://doi.org/10.1111/j.1365-2966.2009.15812.x}{ {\em Mon. Not. Roy. Astron. Soc.} 401 (2010) 2148-2168}
\newblock {[\href{https://arxiv.org/abs/0907.1660}{0907.1660}]}.





\bibitem{Perlmutter:1998np}
S.~Perlmutter et~al.,
\newblock {\em Measurements of $\Omega$ and $\Lambda$ from 42 high redshift
  supernovae},
\newblock \href{ https://doi.org/10.1086/307221}{ {\em Astrophys. J.} 517 (1999) 565-586}
\newblock {[\href{https://arxiv.org/abs/astro-ph/9812133}{astro-ph/9812133}]}.



\bibitem{Agrawal:2018own}
Prateek Agrawal, Georges Obied, Paul~J. Steinhardt, and Cumrun Vafa,
\newblock {\em On the Cosmological Implications of the String Swampland},
\newblock \href{ https://doi.org/10.1016/j.physletb.2018.07.040}{ {\em Phys. Lett. B} 784 (2018) 271-276}
\newblock {[\href{https://arxiv.org/abs/1806.09718}{1806.09718}]}.









\bibitem{Achucarro:2018vey}
Ana Ach\'ucarro and Gonzalo~A. Palma,
\newblock {\em The string swampland constraints require multi-field inflation},
\newblock \href{ https://doi.org/10.1088/1475-7516/2019/02/041}{ {\em JCAP} 02 (2019) 041}
\newblock {[\href{https://arxiv.org/abs/1807.04390}{1807.04390}]}.






\bibitem{Dias:2018ngv}
Mafalda Dias, Jonathan Frazer, Ander Retolaza, and Alexander Westphal,
\newblock {\em Primordial Gravitational Waves and the Swampland},
\newblock \href{ https://doi.org/10.1002/prop.201800063}{ {\em Fortsch. Phys.} 67(1-2) (2019) 2}
\newblock {[\href{https://arxiv.org/abs/1807.06579}{1807.06579}]}.




\bibitem{Garg:2018reu}
Sumit~K. Garg and Chethan Krishnan,
\newblock {\em Bounds on Slow Roll and the de Sitter Swampland},
\newblock \href{ https://doi.org/10.1007/JHEP11(2019)075}{ {\em JHEP} 11 (2019) 075}
\newblock {[\href{https://arxiv.org/abs/1807.05193}{1807.05193}]}.




\bibitem{Kehagias:2018uem}
Alex Kehagias and Antonio Riotto,
\newblock {\em A note on Inflation and the Swampland},
\newblock \href{ https://doi.org/10.1002/prop.201800052}{ {\em Fortsch. Phys.} 66(10) (2018) 1800052}
\newblock {[\href{https://arxiv.org/abs/1807.05445}{1807.05445}]}.




\bibitem{Kinney:2018nny}
William~H. Kinney, Sunny Vagnozzi, and Luca Visinelli,
\newblock {\em The zoo plot meets the swampland: mutual (in)consistency of
  single-field inflation, string conjectures, and cosmological data},
\newblock \href{ https://doi.org/10.1088/1361-6382/ab1d87}{{\em Class. Quant. Grav.} 36(11) (2019) 117001}
\newblock {[\href{https://arxiv.org/abs/1808.06424}{1808.06424}]}.




\bibitem{Ferrara:2010in}
Sergio Ferrara, Renata Kallosh, Andrei Linde, Alessio Marrani, and Antoine
  Van~Proeyen,
\newblock {\em Superconformal Symmetry, NMSSM, and Inflation},
\newblock \href{ https://doi.org/10.1103/PhysRevD.83.025008}{{\em Phys. Rev. D} 83 (2011) 025008}
\newblock {[\href{https://arxiv.org/abs/1008.2942}{1008.2942}]}.








\bibitem{Burgess:2001fx}
C.~P. Burgess, M.~Majumdar, D.~Nolte, F.~Quevedo, G.~Rajesh, and Ren-Jie Zhang,
\newblock {\em The Inflationary brane anti-brane universe},
\newblock \href{https://doi.org/10.1088/1126-6708/2001/07/047}{{\em JHEP} 07 (2001) 047}
\newblock \href{https://arxiv.org/abs/hep-th/0105204}{[hep-th/0105204]}.

\bibitem{Dvali:1998pa}
G.~R. Dvali and S.~H.~Henry Tye,
\newblock {\em Brane inflation},
\newblock \href{https://doi.org/10.1016/S0370-2693(99)00132-X}{{\em Phys. Lett. B} 450 (1999) 72--82}
\newblock \href{https://arxiv.org/abs/hep-ph/9812483}{[hep-ph/9812483]}.

\bibitem{Kachru:2003sx}
Shamit Kachru, Renata Kallosh, Andrei~D. Linde, Juan~Martin Maldacena, Liam~P.
  McAllister, and Sandip~P. Trivedi,
\newblock {\em Towards inflation in string theory},
\newblock \href{https://doi.org/10.1088/1475-7516/2003/10/013}{{\em JCAP} 10 (2003) 013}
\newblock \href{https://arxiv.org/abs/hep-th/0308055}{[hep-th/0308055]}.








\bibitem{Kallosh:2007ig}
Renata Kallosh,
\newblock {\em On inflation in string theory},
\newblock \href{https://doi.org/10.1007/978-3-540-74353-8_4}{{\em Lect. Notes Phys.} 738 (2008) 119--156}
\newblock \href{https://arxiv.org/abs/hep-th/0702059}{[hep-th/0702059]}.

\bibitem{Hebecker:2012aw}
Arthur Hebecker, Sebastian~C. Kraus, Moritz Kuntzler, Dieter Lust, and Timo
  Weigand,
\newblock {\em Fluxbranes: Moduli Stabilisation and Inflation},
\newblock \href{https://doi.org/10.1007/JHEP01(2013)095}{{\em JHEP} 01 (2013) 095}
\newblock \href{https://arxiv.org/abs/1207.2766}{[1207.2766]}.




\bibitem{Kallosh:2010xz}
Renata Kallosh, Andrei Linde, and Tomas Rube.
\newblock {\em General inflaton potentials in supergravity},
\newblock \href{https://doi.org/10.1103/PhysRevD.83.043507}{{\em Phys. Rev. D} 83 (2011) 043507}
\newblock {[\href{https://arxiv.org/abs/1011.5945}{1011.5945}]}.






\bibitem{Roest:2013aoa}
Diederik Roest, Marco Scalisi, and Ivonne Zavala,
\newblock {\em K\"ahler potentials for Planck inflation},
\newblock \href{https://doi.org/10.1088/1475-7516/2013/11/007}{{\em JCAP} 11 (2013) 007}
\newblock \href{https://arxiv.org/abs/1307.4343}{[1307.4343]}.











\bibitem{Dudas:2015eha}
E.~Dudas, S.~Ferrara, A.~Kehagias, and A.~Sagnotti.
\newblock {\em Properties of Nilpotent Supergravity},
\newblock \href{https://doi.org/10.1007/JHEP09(2015)217}{{\em JHEP} 09 (2015) 217}
\newblock \href{https://arxiv.org/abs/1507.07842}{[1507.07842]}.

\bibitem{Ferrara:2014kva}
Sergio Ferrara, Renata Kallosh, and Andrei Linde,
\newblock {\em Cosmology with Nilpotent Superfields},
\newblock \href{https://doi.org/10.1007/JHEP10(2014)143}{{\em JHEP} 10 (2014) 143}
\newblock \href{https://arxiv.org/abs/1408.4096}{[1408.4096]}.

\bibitem{Ferrara:2015tyn}
Sergio Ferrara, Renata Kallosh, and Jesse Thaler,
\newblock {\em Cosmology with orthogonal nilpotent superfields},
\newblock \href{https://doi.org/10.1103/PhysRevD.93.043516}{{\em Phys. Rev. D} 93(4) (2016) 043516}
\newblock \href{https://arxiv.org/abs/1512.00545}{[1512.00545]}.

\bibitem{Kallosh:2014via}
Renata Kallosh and Andrei Linde,
\newblock {\em Inflation and Uplifting with Nilpotent Superfields},
\newblock \href{https://doi.org/10.1088/1475-7516/2015/01/025}{{\em JCAP} 01 (2015) 025}
\newblock \href{https://arxiv.org/abs/1408.5950}{[1408.5950]}.











\bibitem{Aganagic:1996pe}
Mina Aganagic, Costin Popescu, and John~H. Schwarz,
\newblock {\em D-brane actions with local kappa symmetry},
\newblock \href{https://doi.org/10.1016/S0370-2693(96)01643-7}{{\em Phys. Lett. B} 393 (1997) 311--315}
\newblock \href{https://arxiv.org/abs/hep-th/9610249}{[hep-th/9610249]}.

\bibitem{Aganagic:1996nn}
Mina Aganagic, Costin Popescu, and John~H. Schwarz,
\newblock {\em Gauge invariant and gauge fixed D-brane actions},
\newblock \href{https://doi.org/10.1016/S0550-3213(97)00180-6}{{\em Nucl. Phys. B} 495 (1997) 99--126}
\newblock \href{https://arxiv.org/abs/hep-th/9612080}{[hep-th/9612080]}.

\bibitem{Bergshoeff:1997kr}
E.~Bergshoeff, R.~Kallosh, T.~Ortin, and G.~Papadopoulos,
\newblock {\em Kappa symmetry, supersymmetry and intersecting branes},
\newblock \href{https://doi.org/10.1016/S0550-3213(97)00470-7}{{\em Nucl. Phys. B} 502 (1997) 149--169}
\newblock \href{https://arxiv.org/abs/hep-th/9705040}{[hep-th/9705040]}.

\bibitem{Bergshoeff:1996tu}
E.~Bergshoeff and P.~K. Townsend,
\newblock {\em Super D-branes},
\newblock \href{https://doi.org/10.1016/S0550-3213(97)00072-2}{{\em Nucl. Phys. B} 490 (1997) 145--162}
\newblock \href{https://arxiv.org/abs/hep-th/9611173}{[hep-th/9611173]}.

\bibitem{Cederwall:1996ri}
Martin Cederwall, Alexander von Gussich, Bengt E.~W. Nilsson, Per Sundell, and
  Anders Westerberg,
\newblock {\em The Dirichlet super p-branes in ten-dimensional type IIA and IIB
  supergravity},
\newblock \href{https://doi.org/10.1016/S0550-3213(97)00075-8}{{\em Nucl. Phys. B} 490 (1997) 179--201}
\newblock \href{https://arxiv.org/abs/hep-th/9611159}{[hep-th/9611159]}.

\bibitem{Cederwall:1996pv}
Martin Cederwall, Alexander von Gussich, Bengt E.~W. Nilsson, and Anders
  Westerberg,
\newblock {\em The Dirichlet super three-brane in ten-dimensional type IIB
  supergravity},
\newblock \href{https://doi.org/10.1016/S0550-3213(97)00071-0}{{\em Nucl. Phys. B} 490 (1997) 163--178}
\newblock \href{https://arxiv.org/abs/hep-th/9610148}{[hep-th/9610148]}.











\bibitem{Ferrara:2016ntj}
Sergio Ferrara and Antoine Van~Proeyen,
\newblock {\em Mass Formulae for Broken Supersymmetry in Curved Space-Time},
\newblock \href{ https://doi.org/10.1002/prop.201600100}{{\em Fortsch. Phys.} 64(11-12) (2016) 896-902}
\newblock {[\href{https://arxiv.org/abs/1609.08480}{1609.08480}]}.



\bibitem{Ferrara:2019tmu}
Sergio Ferrara, Magnus Tournoy, and Antoine Van~Proeyen,
\newblock {\em de Sitter Conjectures in $N$=1 Supergravity},
\newblock \href{ https://doi.org/10.1002/prop.201900107}{{\em Fortsch. Phys.} 68(2) (2020) 1900107}
\newblock {[\href{https://arxiv.org/abs/1912.06626}{1912.06626}]}.









\bibitem{Lee:2010hj}
Hyun~Min Lee,
\newblock {\em Chaotic inflation in Jordan frame supergravity},
\newblock \href{https://doi.org/10.1088/1475-7516/2010/08/003}{{\em JCAP} 08 (2010) 003}
\newblock \href{https://arxiv.org/abs/1005.2735}{[1005.2735]}.









				


\bibitem{Boran:2019vxz}
Sibel Boran, Emre~Onur Kahya, Nese Ozdemir, Mehmet Ozkan, and Utku Zorba,
\newblock {\em Superconformal generalizations of auxiliary vector modified
  polynomial f(R) theories},
\newblock \href{https://doi.org/10.1088/1475-7516/2020/04/005}{{\em JCAP} 04 (2020) 005}
\newblock {[\href{https://arxiv.org/abs/1912.01919}{1912.01919}]}.


\bibitem{Kehagias:2013mya}
Alex Kehagias, Azadeh Moradinezhad~Dizgah, and Antonio Riotto,
\newblock {\em Remarks on the Starobinsky model of inflation and its descendants},
\newblock \href{https://doi.org/10.1103/PhysRevD.89.043527}{{\em Phys. Rev. D} 89(4) (2014) 043527}
\newblock {[\href{https://arxiv.org/abs/1312.1155}{1312.1155}]}.









\bibitem{Barbon:2009ya}
J.~L.~F. Barbon and J.~R. Espinosa,
\newblock {\em On the Naturalness of Higgs Inflation},
\newblock \href{https://doi.org/10.1103/PhysRevD.79.081302}{{\em Phys. Rev. D} 79 (2009) 081302}
\newblock \href{https://arxiv.org/abs/0903.0355}{[0903.0355]}.

\bibitem{Chakravarty:2013eqa}
Girish Chakravarty, Subhendra Mohanty, and Naveen~K. Singh,
\newblock {\em Higgs Inflation in $f(\Phi,R)$ Theory},
\newblock \href{https://doi.org/10.1142/S0218271814500291}{{\em Int. J. Mod. Phys. D} 23(4) (2014) 1450029}
\newblock \href{https://arxiv.org/abs/1303.3870}{[1303.3870]}.

\bibitem{Kaiser:2013sna}
David~I. Kaiser and Evangelos~I. Sfakianakis,
\newblock {\em Multifield Inflation after Planck: The Case for Nonminimal
  Couplings},
\newblock \href{https://doi.org/10.1103/PhysRevLett.112.011302}{{\em Phys. Rev. Lett.} 112(1) (2014) 011302}
\newblock \href{https://arxiv.org/abs/1304.0363}{[1304.0363]}.

\bibitem{Kallosh:2013tua}
Renata Kallosh, Andrei Linde, and Diederik Roest,
\newblock {\em Universal Attractor for Inflation at Strong Coupling},
\newblock \href{https://doi.org/10.1103/PhysRevLett.112.011303}{{\em Phys. Rev. Lett.} 112(1) (2014) 011303}
\newblock \href{https://arxiv.org/abs/1310.3950}{[1310.3950]}.

\bibitem{Pallis:2010wt}
C.~Pallis,
\newblock {\em Non-Minimally Gravity-Coupled Inflationary Models},
\newblock \href{https://doi.org/10.1016/j.physletb.2010.08.004}{{\em Phys. Lett. B} 692 (2010) 287--296}
\newblock \href{https://arxiv.org/abs/1002.4765}{[1002.4765]}.

\bibitem{Qiu:2012ia}
Taotao Qiu,
\newblock {\em Reconstruction of a Nonminimal Coupling Theory with Scale-invariant
  Power Spectrum},
\newblock \href{https://doi.org/10.1088/1475-7516/2012/06/041}{{\em JCAP} 06 (2012) 041}
\newblock \href{https://arxiv.org/abs/1204.0189}{[1204.0189]}.










\bibitem{Ferrara:2010yw}
Sergio Ferrara, Renata Kallosh, Andrei Linde, Alessio Marrani, and Antoine
  Van~Proeyen,
\newblock {\em Jordan Frame Supergravity and Inflation in NMSSM},
\newblock \href{https://doi.org/10.1103/PhysRevD.82.045003}{{\em Phys. Rev. D} 82 (2010) 045003}
\newblock {[\href{https://arxiv.org/abs/1004.0712}{1004.0712}]}.





\end{thebibliography}
\end{document}